# Poststroke rehabilitative mechanisms in individualized fatigue level-controlled treadmill training- a Rat Model Study

Yuchen Xu[1,2], Yulong Peng[2], Yuanfa Yao[3], Xiaoman Fan[2], Minmin Wang[2,4], Feng Gao[5], Mohamad Sawan[1*], Shaomin Zhang[2*], Xiaoling Hu[6]

[1]CenBRAIN Neurotech Center of Excellence, School of Engineering, Westlake University, Hangzhou, China

[2]Key Laboratory of Biomedical Engineering of Ministry of Education, Qiushi Academy for Advanced Studies, Zhejiang University, Hangzhou, China

[3]The Affiliated Huizhou Hospital, Guangzhou Medical University, Guangzhou, 511436, China

[4]Westlake Institute for Optoelectronics, Westlake University, HangZhou, 310007, China

[5]Department of Neurology, The Second Affiliated Hospital, School of Medicine, Zhejiang University, Hangzhou, 310009, Zhejiang, China

[6]Department of Biomedical Engineering, The Hong Kong Polytechnic University, Kowloon, Hong Kong SAR, China

## Abstract

Individualized training improved post-stroke motor function rehabilitation efficiency. However, there remains limited clarity regarding the mechanisms of how individualized training facilitates recovery. This study aims to explore the cortical and corticomuscular rehabilitative effects in post-stroke motor function recovery during individualized training. In this study, Sprague-Dawley rats with intracerebral hemorrhage (ICH) were randomly distributed into two groups: forced training (FOR-T, n=13) and individualized fatigue-controlled training (FAT-C, n=13) to receive training respectively from day 2 to day 14 post-stroke. Behavior score was evaluated through modified neurological severity score (mNSS), peripheral fatigue was evaluated by electromyography (EMG) mean power frequency (MPF), central fatigue was evaluated through plasma (tryptophan and tyrosine ratio, Trp/Tyr) and electroencephalograph (EEG) Lempel-ziv complexity (LZC), the rehabilitative effects in inter-hemisphere balance and corticomuscular pathways were evaluated by EEG power spectral density (PSD) slope and directed corticmuscular coherence (dCMC) respectively. The FAT-C group exhibited superior motor function recovery (2-way ANOVA, timepoint: $P = 0.000$, $EF = 0.614$, group: $P = 0.000$, $EF = 0.232$) and less central fatigue (Trp/Tyr, Paired t test, FOR-T: $P = 0.0184$, FAT-C: $P = 0.1028$) compared to the FOR-T group. EEG PSD slope analysis demonstrated a better inter-hemispheric balance in FAT-C group compare to the FOR-T group. The dCMC analysis indicated that training-induced fatigue led to a short-term down-regulation of descending corticomuscular coherence (dCMC) and an up-regulation of ascending dCMC. In the long term, excessive fatigue hindered the recovery of descending control in the affected hemisphere. The individualized strategy of peripheral fatigue-controlled training achieved better motor function recovery, which could be attributed to the mitigation of central fatigue, optimization of inter-hemispheric balance and enhancement of descending control in the affected hemisphere.

**Graphical Abstract**

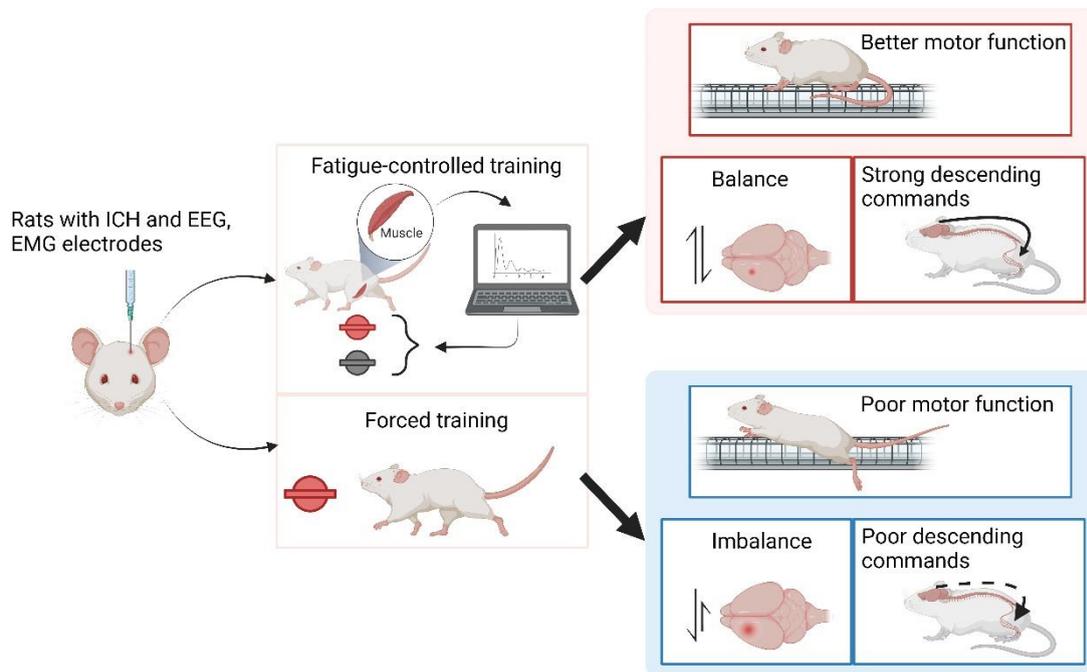

## 1. Introduction

Stroke is the leading cause of chronic disability in adults, with more than 50% of survivors experiencing moderate to severe motor function deficits even after routine rehabilitation (Dimyan and Cohen 2011; Hendricks et al. 2002; Langhorne, Coupar, and Pollock 2009). Early rehabilitation interventions are widely recognized for enhancing recovery outcomes (Wang et al. 2022; Wei et al. 2024). However, excessive fixed-training during the early stages can lead to fatigue, stress and the release of inflammatory cytokines, which may exacerbate secondary brain damage and hinder recovery (Coleman et al. 2017; Langhorne et al. 2017; Li et al. 2017; Sun et al. 2014). Individualized rehabilitation, particularly those incorporating fatigue level modulation through EMG and fatigue-adaptive training, offer a promising approach to optimize training intensity during the early stages of rehabilitation. These tailored strategies have been associated with enhanced neuronal plasticity, improved motor function recovery, reduced stress, and decreased infarct volume compared to conventional fixed-intensity training(Chen, Wang, and Chang 2019; Sun et al. 2014). Despite these advancements, a significant gap remains in understanding how fatigue-constrained, individualized training intensity provides superior benefits for motor function recovery compared to fixed-intensity training that induces excessive fatigue.

Fatigue is a critical biomarker for evaluating the suitability of rehabilitation training intensity. Overloaded training-induced fatigue has been shown to hinder functional recovery while low-intensity training without fatigue often yields poor recovery outcomes (Ballester et al. 2022; Joshi and Dando 2024; Mahrukh et al. 2023; Zeng et al. 2024). Training-induced fatigue occurs in both the central nervous system (central fatigue) and peripheral nervous system (peripheral fatigue). Central fatigue defined as a reduction in voluntary activation of muscles, often associated with a decreased frequency and synchronization of motor neuron activity, and a diminished drive from motor cortex (Tornero-Aguilera et al. 2022). Peripheral fatigue, on the other hand, involves a decline in contractile strength of muscle fibers, often manifested through reduced muscle fiber conduction velocity. Both central and peripheral fatigue are regulated by dynamic negative feedback activity to maintain homeostasis. Peripheral fatigue triggers inhibitory feedback from group III/IV muscle afferents, which suppress neural command output, contributing to central fatigue and mitigating further peripheral fatigue (Amann et al. 2013). Central fatigue reduces motor cortex excitability (Brasil-Neto, Cohen, and Hallett 1994; Deshpande and Hu 2011; McKay et al. 1995) and is associated with increased serotonin levels and decreased dopamine concentrations (Meeusen et al. 2006), potentially impaired cortical neuroplasticity, which is essential to post-stroke rehabilitation (Persico et al. 2001;

D. Wang et al. 2020). Although peripheral fatigue is frequently used as a metric for rehabilitation monitoring, the evaluation of central fatigue in stroke rehabilitation remains scarce. It is unclear whether rehabilitation strategies targeting peripheral fatigue are effective in mitigating the development of central fatigue, underscoring the need for further research in this area.

Post-stroke movement-related rehabilitation effects are predominantly observed at both cortical and corticomuscular levels. At the cortical level, stroke disrupts the interhemispheric balance, leading to increased power in lower frequency bands (delta, theta) and decreased power in higher frequency bands (alpha, beta) in the affected hemisphere (Cassidy et al. 2020; Finnigan, Wong, and Read 2016; Van Wijngaarden et al. 2016). During rehabilitation, the restoration of interhemispheric balance and shifts in power bands in the affected hemisphere have been documented and these changes are considered as biomarkers indicative of motor recovery capacity (Casula et al. 2021; Lefaucheur et al. 2020). At corticomuscular level, functional connectivity during movement execution between cortex and muscle in descending and ascending pathways can be elucidated through directed corticomuscular coherence (dCMC) (Liu, Sheng, and Liu 2019). Studies have reported the up-regulation of descending dCMC was the main reason for the motor function recovery (Khademi et al. 2022) and an augmentation in ascending dCMC to compensate the reduction in descending dCMC (Zhou et al. 2021). These findings underscore the necessity of evaluating rehabilitation effects on both cortical and corticomuscular levels to comprehend the functional mechanisms underlying various training strategies.

This study aims to investigate how peripheral fatigue-based individualized training interacts with central fatigue and induces rehabilitative effects in cortical and corticomuscular levels to achieve motor function recovery. Such exploration is anticipated to contribute to a deeper understanding of the advantages associated with individualized rehabilitation training.

## 2  Materials and Methods

In this study, rats with intracerebral hemorrhage (ICH) were categorized into two groups: the forced training (FOR-T) and fatigue-controlled training (FAT-C) groups, where they underwent repsective rehabilitation interventions from subacute to early chronic periods. During rehabilitation, the central fatigue level, rehabilitative effects at cortical and corticomuscualr levels were investigated and compared between the two groups to elucidate how individualized fatigue-controlled training contributes to motor function recovery in contrast to fixed intensity training. All animal procedures and protocols were approved by the Animal Care Committee of Zhejiang University (protocol number ZJU20210196).

**2.1 Training interventions**

FAT-C and FOR-T training were administered as individualized fatigue-controlled and fixed intensity training, utilizing a custom treadmill training platform (Xu et al. 2020). Both training modes involved an accumulated running duration of 30 minutes per day at a speed of 16m/min. In the FOR-T mode, rats were compelled to continuously run on the treadmill without rest throughout a single trial. Conversely, in the FAT-C mode, the peripheral fatigue level of the affected (AFF) hindlimb was regulated by pausing the training for rest (3 minutes) or restarting it when the fatigue level exceeded or dropped below a preset threshold. The fatigue level was monitored using the drop rate of EMG mean power frequency (MPF). Briefly,

$$MPF\ drop\ rate\ =\ \frac{MPF_{baseline} - MPF_{running}}{MPF_{baseline}} \cdot 100\%\ (1)$$

$$MPF = \frac{\int_{60}^{200} f \cdot s(f)}{\int_{60}^{200} s(f)}\ (2)$$

where $s(f)$ was the power spectrum density of the EMG signal. The MPF value was calculated every 4 seconds and pre-set MPF drop rate threshold was 11 % (Xu et al. 2020). The running MPF was the MPF values calculated based on the EMG of the target hindlimb during treadmill training.

**2.2 Animal Preparation and Surgeries of ICH, EEG and EMG Electrodes Implantation**

Forty-two adult Sprague-Dawley rats (weighing between 270-310g) were included in the study and housed in a vivarium with a 12/12h light/dark cycle, providing ad libitum access to food and water except during experimental periods. An accommodation training phase was conducted over three consecutive days to select rats capable of adapting to continuous treadmill running for subsequent experiments. Thirty-nine rats successfully completed the adaptation training and subsequently underwent surgeries for ICH, as well as the implantation of EEG and EMG electrodes.

In this study, surgeries for ICH, EMG and EEG electrodes implantation were performed sequentially in a single operation. First, ICH and EMG electrodes implantation surgeries were delivered as (Xu et al. 2020). Briefly, ICH was induced in the caudate nucleus (AP: +0.2 mm, ML: +3.0 mm, DV: +6.0mm relative to the Bregma, Type IV collagenase,1.2 μL, 0.25 U in 1μL NaCl 0.9%, C5138, Sigma, USA). For EMG, intramuscular electrodes were implanted in the medial gastrocnemius (MG) muscles of the AFF and unaffected (UN) hindlimbs using a differential electrode configuration (Teflon-coated stainless steel wires; AS632, Cooner Wire, USA) Subsequently, EEG electrodes were implanted. Four stainless steel screw electrodes (diameter: 0.8 mm, length: 1 mm) were threaded into pre-drilled holes in the skull to contact the dura. Each screw electrode was tethered to a silver wire and connected to an 8-pin head connector (Figure 1B). The electrodes were assigned as follows: 1) Motor cortex signal acquisition (Figure 1B, orange dots: AP: -1.75 mm, ML: ±2.25 mm relative to the Bregma), 2) Reference electrode (Figure 1B, green dot: AP: +2 mm relative to the Bregma), 3) Common ground for both EEG and EMG acquisition (Figure 1B, blue dot: AP: -2 mm relative to the Lambda). Additionally, four stainless steel screws (diameter: 1.0 mm, length: 1.2 mm) were inserted into the skull periphery to secure the head connector (locations: AP: +2.0 mm, ML: ±2.0 mm to the Bregma; AP: +2.0 mm, ML: ±3.0 mm to the Lambda). The connector was fixed to these screws with dental cement. After suturing the wounds, the rats were placed in a warm recovery box to maintain body temperature at 36–37°C. Food and water were provided ad libitum until they fully regained consciousness. Six rats were excluded within the first two days post-surgery, primarily due to severe cerebral hemorrhage.

## 2.3 Randomization

After a two-day recovery period following the surgeries, thirty-three rats survived. The quality of both EEG and EMG signals was visually inspected by experimental operators. After a two-day recovery period following the surgeries, thirty-three rats survived. The quality of both EEG and EMG signals was visually inspected by experimental operators before implementing stratified randomization and two rats dropped out due to poor quality of EEG/EMG signals. Additionally, three rats with mild or no motor function deficits (mNSS ≤ 6) were excluded. Twenty-eight rats with moderate to severe motor function deficits (mNSS > 6) on Day2 post-stroke were stratified randomized into two groups, i.e., the FAT-C (n=14) and the FOR-T (n=14) groups. Two rats dropped out during rehabilitation training due to poor quality of EEG/EMG signals. The consort flowchart of the experimental design was shown in Figure 2.

## 2.4 Evaluation

The modified Neurologic Severity Score (mNSS) was employed to assess motor function impairments. Central fatigue was evaluated by measuring the concentrations of tryptophan (Trp) and tyrosine (Tyr) in plasma, as well as through EEG complexity.

Rehabilitative effects induced by training in cortical and corticomuscular pathways were assessed using the EEG power spectral density (PSD) slope and dCMC. The EEG and EMG signals were amplified with a gain of 1,750 using a neural signal acquisition system (OmniPlex Neural Signal Acquisition System, Plexon Inc., USA), with a sampling rate of 40 kHz. Recordings were taken both before and during daily training, representing the awake and training states, respectively. EEG and EMG data were notch filtered between 49 and 51 Hz to remove artifacts and line noise, and band-pass filtered from 2 to 200 Hz during the evaluation phase. Visual inspection of the EEG and EMG signals was conducted to reject any segments or channels with motion artifacts. A total of 364 trials were used in this study.

### 2.4.1    Modified Neurologic Severity Score

Neurological function was graded by mNSS on a scale of 0–18, with 0 indicating normal neurologic function and 18 representing the maximum functional deficits (Schaar, Brenneman, and Savitz 2010; Zhang, Liu, and Liu 2024). The mNSS included assessments of motor function (0-6), sensory (0-2), beam balance (0-6), and reflexes/abnormal movements (0-4). Assessments were conducted by an experimenter who was blinded to the training protocol and group information of the rats from Day2 to Day14, prior to the daily rehabilitation intervention.

### 2.4.2    The concentration of tryptophan and tyrosine in plasma

The blood sample was collected before and after training session to assess the central fatigue induced by the FAT-C and FOR-T training. To mitigate the impact of hemostasis on health status and behavioral evaluation, blood collection was performed on Day 14. After anesthesia with 5% isoflurane in oxygen (2.4l/min), blood was collected from the tail vein within 3min to prevent recovery from central fatigue, ensuring that biomarker levels in the blood were not influenced. Approximate 1 to 2ml of blood sample was collected in plastic tubes containing EDTA- 2Na (7ml, Solarbio, China). The blood-collecting tubes were inverted several times to prevent hemolysis and centrifuged at 2010 g for 10 min at 4°C to obtain plasma. The plasma samples were stored at −80°C until analysis (KARAKAWA et al. 2019). For analysis, 0.1ml of the plasma samples were extracted with 0.2ml acetonitrile and 0.3ml 0.2M hydrochloric acid for 1 h with gentle agitation on a shaker at room temperature. Each sample was filtered through a 0.22 μm pore membrane filter. Then 10 μL of sample was taken into a UHPLC vial and added 70 μL Borate buffer and 2 μL AccQ-Tag reagent. The reaction mixture was kept at room temperature for 1 min, heated at 55 ºC for 10 min and 1 μL was injected after cooling. The sample extracts were analyzed using an UPLC– Orbitrap-MS system (UPLC, Vanquish; MS, QE). Data were acquired on the Q-Exactive using Xcalibur 4.1 (Thermo Scientific), and processed using TraceFinder 4.1 Clinical (Thermo Scientific). The change rate of central fatigue related amino acids was calculated as

$$Change\ rate = \frac{Ratio_{before} - Ratio_{after}}{Ratio_{before}} \quad (3)$$

Where the $Ratio_{before}$ could be the Trp/Tyr, Trp/BCAAs, Tyr/BCAAs acquired from the blood collected before training and the $Ratio_{after}$ could be the Trp/Tyr, Trp/BCAAs, Tyr/BCAAs acquired from the blood collected after training.

### 2.4.3 Lempel-Ziv complexity in EEG

We utilized the drop rate of LZC as a measure of central fatigue level in motor cortex. The LZC calculates the number of new sub-sequence in one segment of the EEG signal. The detailed method is described as (Aboy et al. 2006). The pre-processed EEG signal during treadmill training was segmented into 5s epochs to calculate the LZC value and the LZC drop rate was evaluated as,

$$LZC\ Drop\ rate = \frac{LZC_{begin} - LZC_{end}}{LZC_{begin}} \quad (4)$$

where the $LZC_{begin}$ represented the average LZC values in the first 2 minutes in the 30-min treadmill training. The $LZC_{end}$ was the average LZC values calculated in the last 2 minutes in the 30-min treadmill training.

### 2.4.4 EEG PSD slope in the rest state

The EEG PSD slope was adopted to evaluate stroke-related EEG alterations in wide-band power and the recovery in the affected hemisphere. The EEG PSD slope was calculated as (Lanzone et al. 2022; Leemburg et al. 2018). Briefly, the pre-processed EEG in the resting state (3min) was segmented into 5-second intervals. The power spectra (2-45Hz) of each segment were log-transformed for each frequency bin (1Hz), and a linear curve was fitted to the log-transformed data to estimate the PSD slope of EEG. To minimize error, the mean EEG PSD slope of all the segments was taken as the EEG PSD slope in the resting state.

The symmetry of PSD slope ($PSDslopeSI$) in the AFF and UN in the resting state evaluated the asymmetry of the two hemispheres as

$$PSDslopeSI = \frac{PSD_{slopeAFF} - PSD_{slopeUN}}{PSD_{slopeAFF} + PSD_{slopeUN}} \quad (5)$$

Where the $PSD_{slopeAFF}$ or the $PSD_{slopeUN}$ represented the PSD slope value in the AFF or the UN hemisphere in the resting state.

### 2.4.5 Directed cortico-muscular coherence

DCMC reflected the interaction between the descending and ascending pathways between the motor cortex and peripheral muscles. We used the dCMC to investigate the interaction changes during the FOR-T and FAT-C post-stroke training. The dCMC was estimated by autoregressive (AR) model as previous studies (Faes and Nollo 2011; Zhou et al. 2021). Briefly, the EEG($X_1(t)$) and EMG($X_2(t)$) data were segmented into 1-s length and depicted as an AR model:

$$X(t) = \sum_{k=0}^{p} X(t-k) + E(t) \quad (6)$$

Where $A_k$ is a 2 × 2 matrix of coefficients describing the causal influence of the signals at lag k on the signals at lag zero, p represents the order of AR model, and $E(t)$ indicates the prediction errors. Equation (6) was transformed to the frequency domain to obtain the transfer function $H(f)$,

$$X(f) = A^{-1}(f)E(f) = H(f)E(f) \quad (7)$$

$H_{12}(f)$ represent the transfer functions giving the influence of EMG on EEG. The directed coherence from EMG to EEG is defined as

$$DC_{12}(f) = \frac{\sigma_2 H_{12}(f)}{\sqrt{\sigma_1^2 |H_{11}(f)|^2 + \sigma_2^2 |H_{12}(f)|^2}} \quad (8)$$

Where $\sigma_1$ and $\sigma_2$ represents the variance of prediction errors in EEG and EMG respectively.

The significant level was estimated by numerical Monte Carlo Simulation. Briefly, two independent Gaussian random vectors with the same number of trials as the original data were generated and the directed coherence was calculated. This procedure was repeated 50 times with different random numbers. Directed coherence estimates at all frequencies and for all simulations were rank ordered and the 95th percentile was used as an approximate P <0.05 significance level. The non-significant dCMC with value lower than the significant level was set to 0 for subsequent statistical comparisons (Zhou et al. 2021). The dCMC amplitude was normalized and the mean dCMC in the first 2 minutes and last 2 minutes during one running trial were compared to demonstrate the effects of FOR-T and FAT-C training.

## 2.5 Statistical Analysis

The normality test was performed by the Kolmogorov-Smirnov test with a significant level of 0.05. All the data samples conformed to normal distributions (P< 0.05). Differences of the MPF drop rates, the behavior scores, the LZC drop rates, the EEG PSD slope, and the baseline dCMC between the FOR-T and FAT-C groups were evaluated using a two-way analysis of variance (ANOVA). The intragroup changes at different timepoints and intergroup differences at the same timepoint were evaluated using one-way ANOVA followed by post-hoc Bonferroni tests. The intragroup changes of free Trp/BCAAs on Day14 before and after training and the changes of dCMC amplitude at the beginning and end of single running trial in the FOR-T and the FAT-C groups were compared through paired t-test. The intergroup comparison of change rate of free Trp/BCAAs on Day14 before and after training and the dominant pattern of the dCMC in the descending and ascending pathways were evaluated by t-test. $P < 0.05$ was adopted as a statistically significant level in this study. Significant levels were indicated as, 1 superscript for < 0.05, 2 superscripts for < 0.01, and 3 superscripts for < 0.001. All statistical analyses were performed using SPSS (Version 20, IBM, USA).

## 3 Results

### 3.1 Peripheral fatigue constrained during FAT-C training

To verify the peripheral fatigue-controlled training strategy, the MPF drop rates during training in the AFF and UN hindlimbs of the FAT-C and FOR-T groups were evaluated from Day2 to Day14 post-stroke. For the AFF hindlimb, the FOR-T group showed a higher MPF drop rate compared the FAT-C group from Day2 to Day11, excluding Day 8 and Day10 (One-way ANOVA, $P \leq 0.002$, $EF \geq 0.282$, Bonferroni's post-hoc: $P \leq 0.026$, Figure 3A). Regarding the UN side, the FOR-T group showed a higher MPF drop rate compared to the FAT-C group from Day2 to Day12, excepted Day4, Day7, Day10 and Day11 (One-way ANOVA, $P \leq 0.006$, $EF \geq 0.283$, Bonferroni's post-hoc: $P \leq 0.035$, Figure 3A). These findings indicated the fatigue level was constrained in the FAT-C group compared to the FOR-T group, and validated the effectiveness of the fatigue-controlled training platform. This observation aligns with the results of our previous study (Xu et al. 2020).

### 3.2 The fatigue-controlled training strategy constrained the peripheral fatigue level during training and led to better functional outcomes compared to forced training

The overall mNSS and subscore in the FOR-T and the FAT-C groups were assessed from Day2 to Day14 post-stroke. The FAT-C group showed significantly lower mNSS compared to the FOR-T group form Day4 to Day14 (1-way ANOVAs, P ≤ 0.045, EF ≥ 0.120, Figure 3B). For the intra-group comparison, the FAT-C group showed significant reduction of mNSS compared to baseline since Day4 while the significant reduction was found since Day5 in the FOR-T group (FAT-C: 1-way ANOVA, P = 0.000, EF = 0.680, Bonferroni's post-hoc tests P ≤0.002; FOR-T: 1-way ANOVA, P=0.000, EF=0.532, Bonferroni's post-hoc tests P ≤ 0.003, Figure 3B). Both the FOR-T and the FAT-C group showed motor function recovery but a better motor function recovery was found in the FAT-C group (Figure 3B).

Furthermore, FAT-C group displayed a significantly lower beam balance sub-score compared to the FOR-T group from Day5 to Day14 (except Day7, 1-way ANOVAs, P ≤ 0.030, EF ≥ 0.138, Figure S1). Additionally, the FAT-C group exhibited a significantly lower motor sub-score compared to the FOR-T group in Day4, Day8 and Day10 to Day14 (1-way ANOVAs, P ≤ 0.029, EF ≥ 0.141, Figure S1). The inter-group differences were concentrated more in the beam balance sub-score compared to the motor subscore. These results demonstrated the both faster and superior motor function recovery in individualized fatigue-controlled training, especially in the movement balance ability compared to FOR-T group. Further analysis suggested the superior recovery mainly observed in the beam balance sub-score, which suggested an improved coordination and balance ability (Luong et al. 2011) in the FAT-C group compared to the FOR-T group. These results further emphasize the importance of individualized training strategy in early stroke rehabilitation.

### 3.3 Central fatigue was constrained in the FAT-C group

During prolonged physical training, free Trp and Tyr in plasma compete with BCAAs to traverse the blood-brain barrier, thereby leading to the synthesis of serotonin and dopamine, resulting in a feeling of lethargy and a decline in neural drive, ultimately leading to central nervous system (CNS) fatigue (Meeusen and Watson 2007; Soares, Coimbra, and Marubayashi 2007). The ratios of free Trp/BCAAs and Tyr/BCAAs are in alignment with the accumulation of serotonin and dopamine in the brain (Cordeiro et al. 2017; Hasegawa et al. 2008; Teixeira-Coelho et al. 2014), so the changes of free Trp/Tyr before and after training in the FAT-C and FOR-T groups were assessed to evaluate central fatigue.

A significant higher value of free Trp/Tyr was observed after training compared to before training in the FOR-T group, while no significant change was found in the FAT-C group (FOR-T: P = 0.0184, FAT-C: P = 0.1028, Figure 4A). The FOR-T group also showed a higher Change Rate of free Trp/Tyr compared to the FAT-C group (P = 0.0399, Figure 4A). Further, Trp/BCAAs and Tyr/BCAAs were analyzed to investigate the concentration of 5-HT and dopamine in brain tissue.

A significant higher value of free Trp/BCAAs was found after training compared to before training in the FOR-T group, while no significant change was found in the FAT-C group (FOR-T: P = 0.0015, FAT-C: P = 0.1888, Figure 4B). For the intergroup comparison, the FOR-T group showed a higher Change RatefreeT rp/BCAAs compared to the FAT- C group (P = 0.0016, Figure 4B). For the changes of Tyr, no significant difference was found between the after training and the before training in the FAT-C and FOR- T groups (FAT-C: P = 0.1320, FOR-T: P = 0.3746, Figure 4C). However, the Change Rate of Tyr/BCAAs showed significant difference between the FAT-C and the FOR-T groups (P = 0.0181, Figure 4C). These results revealed significant central fatigue related neurotransmitter generated during forced training and mainly be attributed to the 5-HT increasement while FAT-C training constrained synthesis of serotonin. Intense release of serotonin could reduce motoneurone activity and exacerbate central fatigue during prolonged sustained contractions (Kavanagh, McFarland, and Taylor 2019). The 5-HT is related to plasticity injury, which may constrain the recovery of the affected hemisphere, further led to the imbalance between the two hemispheres. Besides, exercise-induced fatigue impaired the corticostriatal long-term potentiation and long-term depression were both impaired (Ma et al. 2018), which were essential in motor function rehabilitation. So, we proposed the individualized training could constrain central fatigue through constraining the peripheral fatigue level, while in the FOR-T group, central fatigue related motor drive reduction and plasticity injury led to the poorer motor function recovery.

The LZC drop rates of the AFF/UN hemisphere in the FOR-T and FAT-C groups from Day2 to Day14 post-stroke were investigated to reveal the central fatigue in motor cortex level. In the AFF hemisphere, significant higher LZC drop rates were found in FOR-T group compared to the FAT-C group from Day2 to Day5 (1- way ANOVA, P ≤ 0.005, EF ≥ 0.259, Bonferroni post-hoc: P ≤ 0.030, Figure 5). For the UN hemisphere, significant higher LZC drop rates were found in the FOR-T group compared to the FAT-C group in Day4 (1-way ANOVA, P = 0.001, EF = 0.311, Bonferroni post- hoc: P = 0.017, Figure 5).

In the late rehabilitation stage (Day14), although no significant difference was observed in motor drive output, as quantified by EEG complexity, the levels of central fatigue-related amino acid in plasma indicated greater central fatigue in the FOR-T group. These findings demonstrated that the peripheral fatigue-controlled strategy can also mitigate central fatigue in late rehabilitation stage and underscore the importance of individualized fatigue-monitoring training during the late stages of rehabilitation.

### 3.4 Forced training aggravated the asymmetry of two hemispheres by strengthening the unaffected hemisphere

The EEG PSD slope of the UN/AFF hemispheres in the FAT-C and the FOR-T groups from Day2 to Day14 post-stroke were investigated to evaluate the rehabilitation effects in hemispheres activity. The EEG PSD slope reduction in the AFF side demonstrated the stroke-induced injury and reflection of increased power in lower frequency bands and decreased power in higher frequency bands. On the late stage of rehabilitation, significant higher EEG PSD slopes were found in the FOR-T UN group compared to the FAT-C UN group from Day11 to Day14 (one-way ANOVAs, $P \leq 0.002$, $EF \leq 0.298$, Bonferroni's post hoc, $P \leq 0.023$, Figure 6A). These results indicated the FOR-T strategy may contribute to strengthen of the UN side, which meanwhile constrain the recovery of the AFF side. The EEG PSD slope SI of the FAT-C and the FOR-T groups from Day2 to Day14 post-stroke were adopted to evaluate inter-hemispheric balance. Significant higher EEG PSD slope SI values were found in the FAT-C group compared to the FOR-T group on Day3, Day4 and Day11 to Day14 (one-way ANOVAs, $P \leq 0.043$, $EF \leq 0.181$, Figure 6B). These results indicated the poorer inter-hemisphere im-balance in the FOR-T group mainly be related to the strengthen of UN side. The imbalance may further constrain the motor function recovery.

### 3.5 Fatigue deteriorated the causal influence from the motor cortex to the target muscle

The normalized descending and ascending dCMC in the early and the late stage of one single treadmill training trial from Day2 to Day14 were evaluated to investigate the influence of fatigue training in corticomuscular pathways.

In descending pathway, both the beta band and the gamma band dCMC was significantly lower in the late stage compared to the early stage from in the FOR-T UN group (Beta band, Day2 to Day6, $P \leq 0.0344$, Figure 7A; Gamma band, Day2 to Day7, $P \leq 0.0496$, Figure 7B ), while no significant difference between the early and the late stage was found in the FOR-T AFF, FAT-C UN, and FAT-C AFF groups ($P > 0.05$, Figure 7C,D); In the AFF side, no significant difference between the early and the late stage was found in both the FOR-T AFF and the FAT-C AFF groups (P> 0.05, Figure 7E-H). These results suggested the motor cortex would decrease descending motor drives in response of extreme fatigue state, to keep a dynamic homeostasis state to avoid injury.

In ascending pathway, both beta and gamma band dCMC were significantly lower in the late stage compared to the early stage from in the FOR-T UN and FOR-T AFF groups (Beta band, FOR-T UN, Day2 to Day9, P ≤ 0.0480, Figure 7I; Gamma band, FOR-T UN, Day2 to Day12 excepted Day7, Day11, P ≤ 0.0491, Figure 7J; Beta band, FOR-T AFF, Day2 to Day10 excepted Day3, Day7, P ≤ 0.0438, Figure 7M; Gamma band, FOR-T AFF, Day2 to Day10, P ≤ 0.0479, Figure 7N). No significant difference between the early and the late stage was found in the FAT-C UN and FAT-C AFF groups (P > 0.05, Figure 7K,L,O,P). These results suggested in the extreme fatigue state with motor cortex decreased the drives to muscle, muscle would increase the ascending feedback to compensate and complete stable movement.

### 3.6 DCMC dominant pattern during rehabilitation

The dCMC dominant main effect in the beta band was shown in Figure 8A-D. For the UN side, significant higher descending dCMC compared to the ascending dCMC was found in both the FAT-C UN and the FOR-T UN groups (FAT-C UN group: P< 0.0001, FOR-T UN group: P <0.0001). For the AFF side, there is no significant difference between the descending and the ascending dCMC in the FAT-C AFF group (P > 0.05), while a significant higher ascending dCMC compared to the descending dCMC was found in the FOR-T AFF group (P = 0.0006). These results demonstrated that stroke disrupted the descending dominant main effect in the beta band. For the FOR-T group, training led to a reduction in descending CMC and an increase in ascending CMC. The control of muscles by the single-synaptic corticospinal tract from the cortex was impaired, significantly weakening the descending pathway. As a result, the group could not achieve descending dominant movement control but instead exhibited an ascending dominance phenomenon, further amplifying the dominant pattern difference, reflected as the ascending dominant main effect. In contrast, for the FAT-C AFF group, no significant dominant main effect was observed, suggesting a recovery in descending control ability.

The dCMC dominant main effect in the gamma band was shown in Figure 8E-F. For the UN side, no significant difference was found between the descending and the ascending dCMC in both the FAT-C UN and the FOR-T UN groups (no dominant pattern, FAT-C UN group: P = 0.5548, FOR-T UN group: P= 0.7743). For the AFF side, significant higher ascending dCMC compared to the descending dCMC was found in both the FAT-C AFF group and the FOR-T AFF group (FAT-C AFF group: P = 0.0043, FOR-T AFF group: P= 0.0007). These results suggested no significant difference was found between the FAT-C and FOR-T training groups regarding gamma band dominant main effect.

## 4. Discussion

The low rehabilitation efficiency of post-stroke motor function rehabilitation remains a significant challenge. Our study aimed to investigate how individualized training benefits post-stroke motor function rehabilitation and offer potential solutions for high efficiency rehabilitation. In this study, we compared the rehabilitation effects of individualized fatigue-controlled and forced rehabilitation training in stroke rat model. We found that an individualized peripheral fatigue-controlled training strategy improved post-stroke motor function recovery by limiting central fatigue, achieving better inter-hemispheric balance and enhancing descending movement control. This is the first study to explore how individualized peripheral fatigue-controlled training improves the rehabilitation outcomes in stroke model, providing a promising approach to enhance post-stroke rehabilitation efficiency.

To further investigate the rehabilitative effects of individualized training, analyses at the cortical and corticomuscular pathway levels were conducted. At the cortical level, the rehabilitative effects were found to be related to inter-hemispheric balance, which is disrupted by stroke due to the reduced inhibitory projections from the affected hemisphere to the unaffected hemisphere (Chen et al. 2023; Dodd, Nair, and Prabhakaran 2017; Mauro et al. 2024). In this study, we found stroke disrupted the inter-hemispheric symmetry of EEG PSD slope on Day2, while a significantly better symmetry index was observed in the FAT-C group during the late stage of rehabilitation (Figure 6). This suggested fatigue-controlled training can alleviate stroke-induced inter-hemispheric imbalance, bringing it to a more balanced state compared to the forced training strategy. The difference in inter-hemispheric imbalance may be related to the varying fatigue levels caused by different training strategies. Peripheral fatigue and central fatigue maintained dynamic homeostasis through negative feedback (Myers 2024; Tornero-Aguilera et al. 2022; Wan et al. 2017). As peripheral fatigue increases during intensive training, it generates central fatigue, further constraining motor cortex output. This aligns with the significantly higher peripheral fatigue and central fatigue observed in the FOR-T group, as evaluated by LZC drop rate and plasma. A reduction in EEG complexity is considered a potential indicator of decreased motor drive commands, attributed to central fatigue (Ibáñez-Molina et al.

2018; Jun Hong 2012; Xu et al. 2018). This reduction in motor drive is consistent with findings of exhaustion in tasks such as pull-ups (Tergau et al. 2000) and thumb isometric contractions (Kotan et al. 2015). Excepted the observation in cortical level, we found FOR-T group experienced a reduction in descending CMC and increasement of ascending CMC. The decline in descending dCMC signifies a reduction in the motor cortex's ability to control muscles and a decoupling of signals between peripheral muscles and the motor cortex, which may be linked to the attenuation of external input to motor neurons due to the suppression of motor cortex by muscle afference in group III/IV in response to peripheral fatigue state (Taylor et al. 2016). In a previous study, we found forced training under extreme fatigue levels could potentially exacerbate damage to the AFF striatum (Xu et al. 2020). This damage further reduces the inhibitory projections from the AFF hemisphere, leading to increased excitability in the UN hemisphere, which disrupts inter-hemispheric balance and constrains the recovery of the AFF hemisphere. Simultaneously, the compromised motor function of the AFF hindlimb may lead to greater reliance on the UN hindlimb, further inhibiting recovery of balance ability. These findings suggest that overloaded training intensity exacerbates stroke-induced inter-hemispheric asymmetry, further hindering recovery in the AFF hemisphere in the FOR-T group. This results in poorer motor function rehabilitation compared to the FAT-C training strategy. In healthy state, the motor cortex may increase excitability in fatigue state as a compensatory role to maintain the motor output, while stroke reduced cortical excitability and increased excitability thresholds to stimulation in the AFF motor cortex (Cicinelli et al. 2003; Hogan et al. 2020; Hsu et al. 2023; Kuppuswamy et al. 2015), making it difficult for the AFF hemisphere to increase motor drive.

    Early rehabilitation interventions offer the potential to leverage the critical window of neural plasticity, leading to significant improvements in behavioral and functional outcomes. However, the implementation of individualized and moderate training strategies in clinical settings is often constrained by limited medical resources and personnel. This gap presents a challenge in tailoring rehabilitation plans to meet the specific needs of each patient. Overloaded training can activate the sympathetic-adrenal medullary and hypothalamus-pituitary-adrenal axes, leading to the release of stress and catabolic hormones, as well as inflammatory cytokines (Reddin et al. 2022; Sun et al. 2014; Taylor et al. 2016), which may hinder functional recovery. In this study, a peripheral fatigue-controlled training strategy provides a potential solution for achieving safe early rehabilitation interventions with high efficiency in clinical settings. This approach visualizes an individual's specific muscle fatigue level and uses feedback to adjust the

training strategy. By tailoring the rehabilitation protocol in this way, the strategy minimizes the risks associated with overtraining and enhances the overall efficiency of the rehabilitation process, even with limited medical resources.

This study also highlighted the critical role of inter-hemispheric balance in early-stage rehabilitation and how fatigue-controlled training can facilitate this rebalancing. Stroke disrupts inter-hemispheric balance, leading to reduced excitability in the affected (AFF) hemisphere. This imbalance in motor activation is associated with motor deficits (Graterol Pérez et al. 2022; Kinoshita et al. 2019; Tam et al. 2024) , which can recover with motor restoration (Tang et al. 2015). In our study, we found that the forced training group exhibited poor inter-hemispheric balance, with stronger activation in the unaffected hindlimb. Overloaded training led to excessive use of both limbs to accomplish movements, which caused stress and injury, resulting in reduced plasticity (Sun et al. 2014). This injury restricted the recovery of the AFF hemisphere, which was further constrained by the UN hemisphere due to inter-hemispheric inhibition. In clinical research, repetitive transcranial magnetic stimulation was applied to facilitate inter-hemispheric rebalancing in severe hemiplegic stroke patients, by either suppression of the UN cortex or activation of the AFF cortex, superior motor recovery was found in the UN cortex suppression group (Q. Wang et al. 2020), emphasizing the importance of suppressing the UN hemisphere. However, in chronic severe stroke, changes in UN hemisphere plasticity are associated with better motor function outcomes, which indicated the role of UN hemisphere in motor function recovery (Demirtas-Tatlidede et al. 2015). For clinical stroke rehabilitation, the amount of structural reserve determines whether inter-hemispheric imbalance or UN hemisphere compensation dominates. If structural reserve is high, the interhemispheric balance model can predict recovery better than the compensation model, while with low structural reserve, UN compensation may dominate (Di Pino et al. 2014; Li et al. 2022). Thus, structural measurements may be necessary to further enhance the efficiency of the FAT-C strategy in clinical applications.

Voluntary movement execution requires the coordination of descending motor drive and ascending sensory feedback. An increase in ascending dCMC is associated with the instability of motor control caused by excessive fatigue. Movement execution relies on the constant comparison of ascending sensory feedback with expected feedback to complete motion calibration. However, when muscles are fatigued, the motor cortex's control ability weakens, impairing its response to ascending feedback. This disruption makes it difficult to appropriately respond to upward feedback, challenging effective motion calibration. As a result, the ascending pathway is continuously enhanced. Similar phenomena have been observed in previous studies, where subjects compensated for unstable muscle states by increasing ascending dCMC to maintain exercise output (Zhou et al. 2021). However, in the later stage of the rehabilitation training intervention (days 10 to 14), the enhancement of the ascending pathway was reduced, indicating that with the improvement of motor function and fatigue tolerance, exercise output during the latter phase of single running training tended to stabilize. In this study, we observed distinct dominant patterns in the beta and gamma bands of dCMC. Previous studies have reported that the beta band primarily transmits downward motion control information (Gwin et al. 2011) while the gamma band mainly transmits upward sensory feedback (Liang et al. 2021). In normal/undamaged movement state, by comparing the expected and actual incoming proprioceptive feedback, the motor output instructions are constantly adjusted to complete and stabilize the motor task, showing the phenomenon of descending dominance (Baker 2007). We found that the FOR-T group showed significantly less descending CMC, as indicated by the ascending dominant effect in the beta band, while the FAT-C group did not. This suggests that fatigue-controlled training effectively enhances the descending pathway on the affected side, improving the ability to control the affected hindlimb and preventing the ascending dominant pattern in beta band dCMC. As a result, the FAT-C group experienced better motor function recovery.

Previous studies suggested that cortical involvement is unnecessary for lower limb movement in rodents, as the repetitive basal movements of lower mammals are primarily regulated by subcortical and spinal cord networks (Stuart and Hultborn 2008). However, recent research has shown consistent activation of motor neurons in the lower extremity motor cortex and the hind limbs of rats during ground walking. Additionally, differences in motor cortex activation have been observed across walking paradigms with varying levels of participation and difficulty (Li et al. 2021). In this study, we further demonstrated that cortical participation is required for rats to perform lower extremity running after a stroke.

In conclusion, this study found that individualized peripheral fatigue-controlled training effectively mitigated central fatigue. This approach promoted inter-hemispheric rebalancing and the recovery of motor drive output, facilitating rehabilitation of the affected hemisphere rather than overstrengthening the unaffected hemisphere, as observed with overloaded forced training. The findings provide further insight into how individualized training strategies benefit post-stroke motor function recovery and offer a potential solution for achieving high-efficiency post-stroke motor function rehabilitation.

## 5. Limitations

The current study demonstrated the rehabilitative effects of peripheral fatigue-controlled rehabilitation strategy using a rat stroke model. However, there are some limitations that warrant and future work should be discussed. Firstly, the peripheral fatigue level was monitored solely through EMG signals. Future studies could incorporate additional biomarkers, such as heart rate or percentage of maximal oxygen uptake, to assess general exercise-induced fatigue rather than focusing on a single muscle group (Li 2022). Secondly, the fatigue threshold was fixed during rehabilitation training. It would be valuable to explore whether modulating fatigue tolerance during rehabilitation could lead to better behavioral outcomes. As motor function improves, training intensity tolerance tends to increase (Fabero-Garrido et al. 2022). An adaptive fatigue threshold might further enhance rehabilitation efficiency. Lastly, central fatigue was evaluated through plasma analysis only on Day 14. Future research should investigate whether this single time point sufficiently represents the trends in central fatigue from Day 2 to Day 14, providing a more comprehensive understanding of its progression.

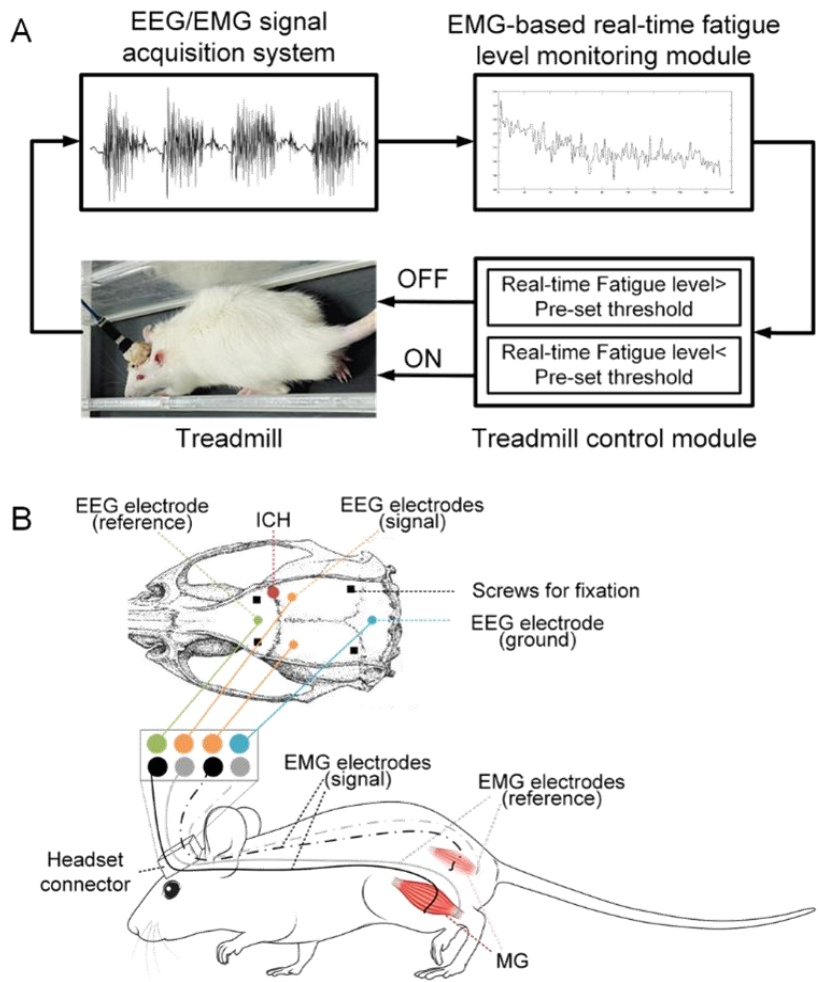

Figure 1 (A) The fatigue-controlled treadmill training platform. The rat was running on the treadmill, its EMG signal was acquired and amplified by EEG/EMG signal acquisition system and transmitted to the EMG-based real-time fatigue level monitoring module. The fatigue level controlled the on/off state of the treadmill. (B) Illustration of the ICH, the EEG electrodes, the headset connector and the EMG electrodes. The dorsal skull surface of the rat (top) showed the locations of the injection hole (red dot) induced ICH, the EEG electrodes for EEG signal in the M1 (orange dots), the EEG electrode for the reference (green dot), the common ground (blue dot), and the screws for fixation (black squares). The rat model (bottom) illustrated the intramuscular electrode implantation on the MG. The black and grey solid/dash lines in the rat model represented the EMG electrodes of the EMG signal (the belly of MG) and the EMG reference (tendon), respectively.

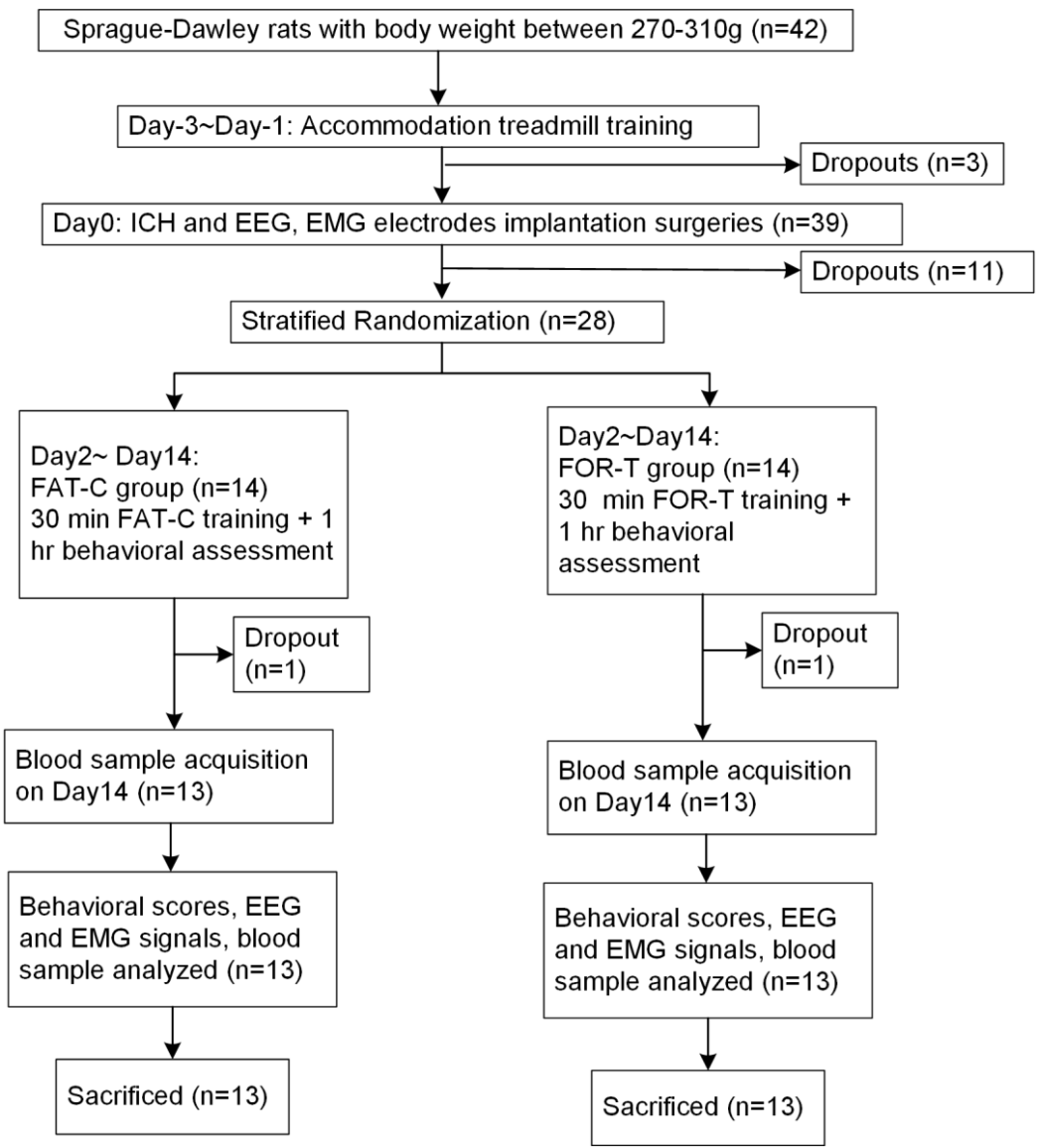

Figure 2 The consort flowchart of the experimental design. All rats received a three-day treadmill accommodation running followed by ICH, EMG and EEG electrodes implantation surgeries. Two days after these operations, the rats were randomly distributed into two groups and received FOR-T or FAT-C from Day2 to Day14 post-stroke. The changes in the central fatigue, inter-hemispheric balance, corticomuscular pathways were evaluated to reveal the mechanisms of individualized peripheral fatigue-based training.

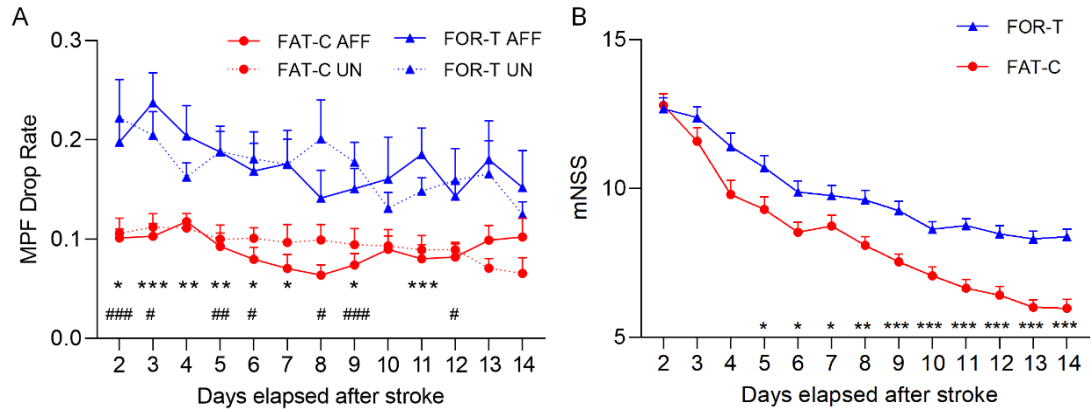

Figure 3 (A) The MPF drop rates of the UNAFF hindlimbs in the FAT-C and FOR-T groups from Day2 to Day14 post-stroke. The "*" and "#" indicated the significant differences between the MPF drop rates of the FAT-C and FOR-T groups in the AFF and the UN hindlimb, respectively. (B) The mNSS of the FAT-C and the FOR-T groups from Day2 to Day14 post-stroke. The "*" i ndicated the significant differences between the FAT-C and FOR-T groups.The values were represented by mean and SEM at each timepoint.

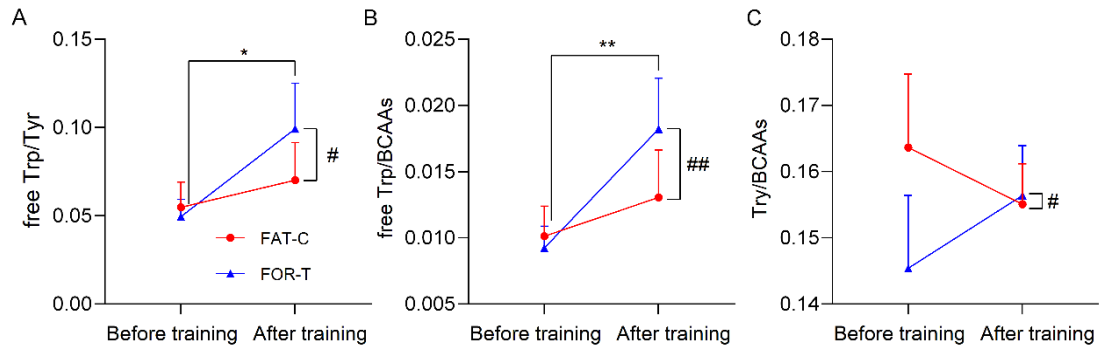

Figure 4 The values of (A) free Trp/Tyr ,(B)free Trp/BCAAs, (C)Tyr/BCAAs before training and after training in the FAT-C and FOR-T groups in Day14. The "*" indicated the significant differences before and after training. The "#" indicated the significant differences between the change rate of amino acids in the FAT-C and the FOR-T groups

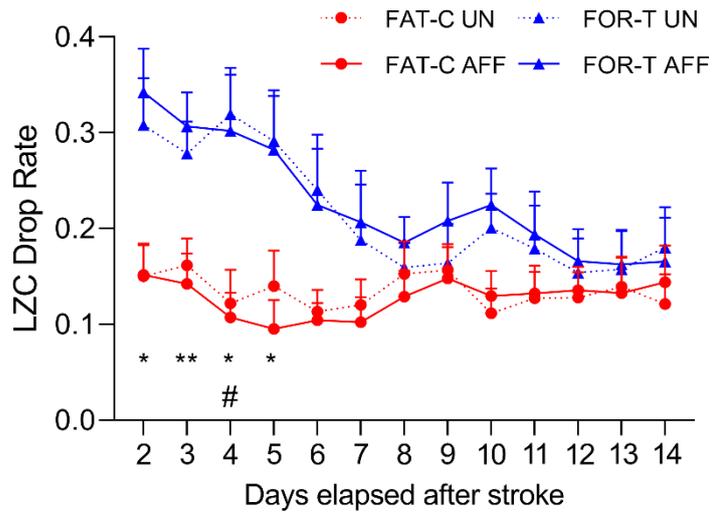

Figure 5 The LZC drop rates of the AFF/UN hemisphere in the FOR-T and FAT-C groups from Day2 to Day14 post-stroke. The values were represented by mean and SEM at each timepoint. The significant differences between the LZC drop rates in the AFF hemisphere of FAT-C and FOR-T groups were indicated by "*". The significant differences between the LZC drop rates in the UN hemisphere of FAT-C and FOR-T groups were indicated by "#".

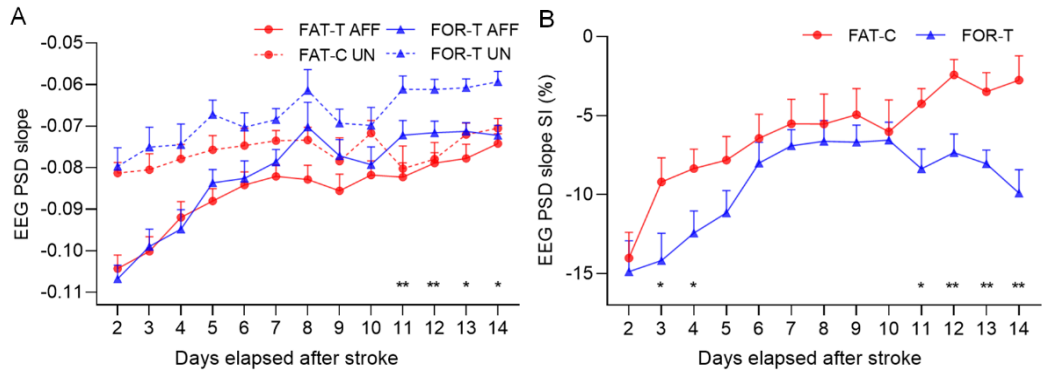

Figure 6 (A). The EEG PSD slope in the AFF/UN hemispheres of the FOR-T and FAT-C groups from Day2 to Day14 post-stroke. The significant differences between the UN hemispheres of the FAT- C and FOR-T groups were indicated by "*". (B) The EEG PSD slope SI in the FOR-T and FAT-C groups from Day2 to Day14 post-stroke. The significant differences between the EEG PSD slope SI of the FAT-C and FOR-T groups were indicated by "*". The values were represented by mean and SEM at each timepoint.

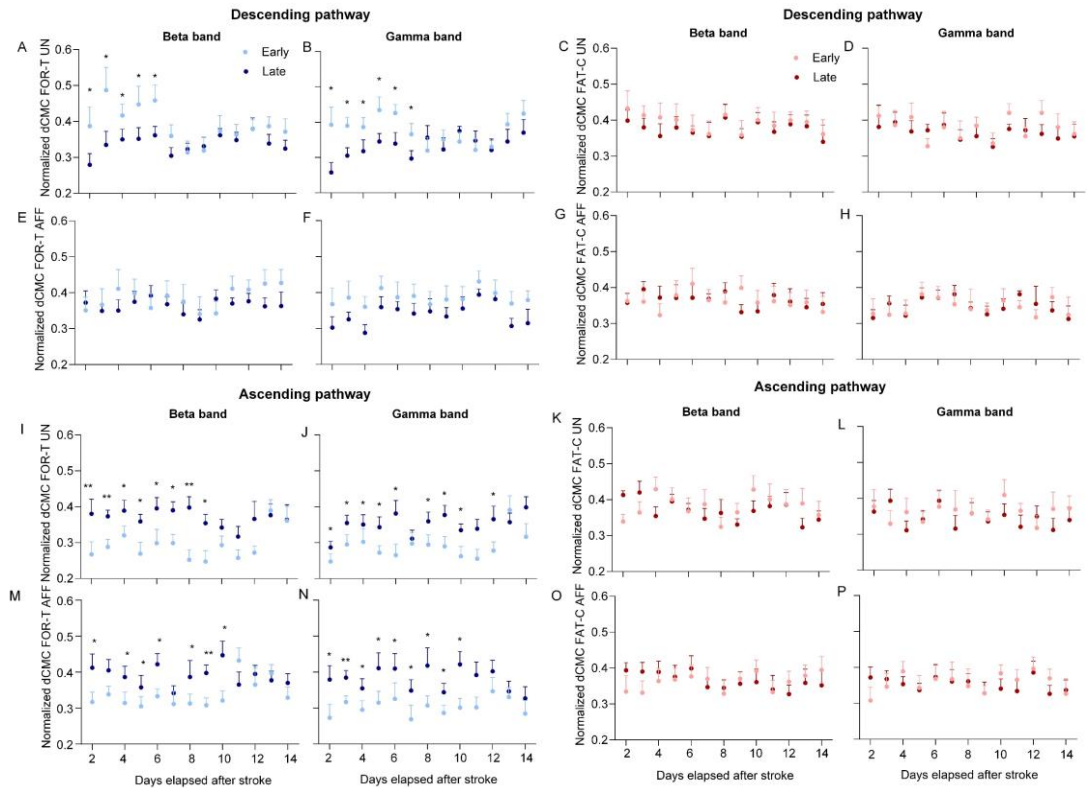

Figure 7 The normalized dCMC in early and late stage of running duration from Day2 to Day14 post-stroke in the FOR-T group. (A-D) illustrated the normalized descending dCMC in beta and gamma band of the FOR-T and FAT-C UN group. (E-H) illustrated the normalized descending dCMC in beta and gamma band of the FOR-T and FAT-C AFF group. (I-L) illustrated the normalized ascending dCMC in beta and gamma band of the FOR-T and FAT-C UN group. (M-P) illustrated the normalized ascending dCMC in beta and gamma band of the FOR-T and FAT-C AFF group.

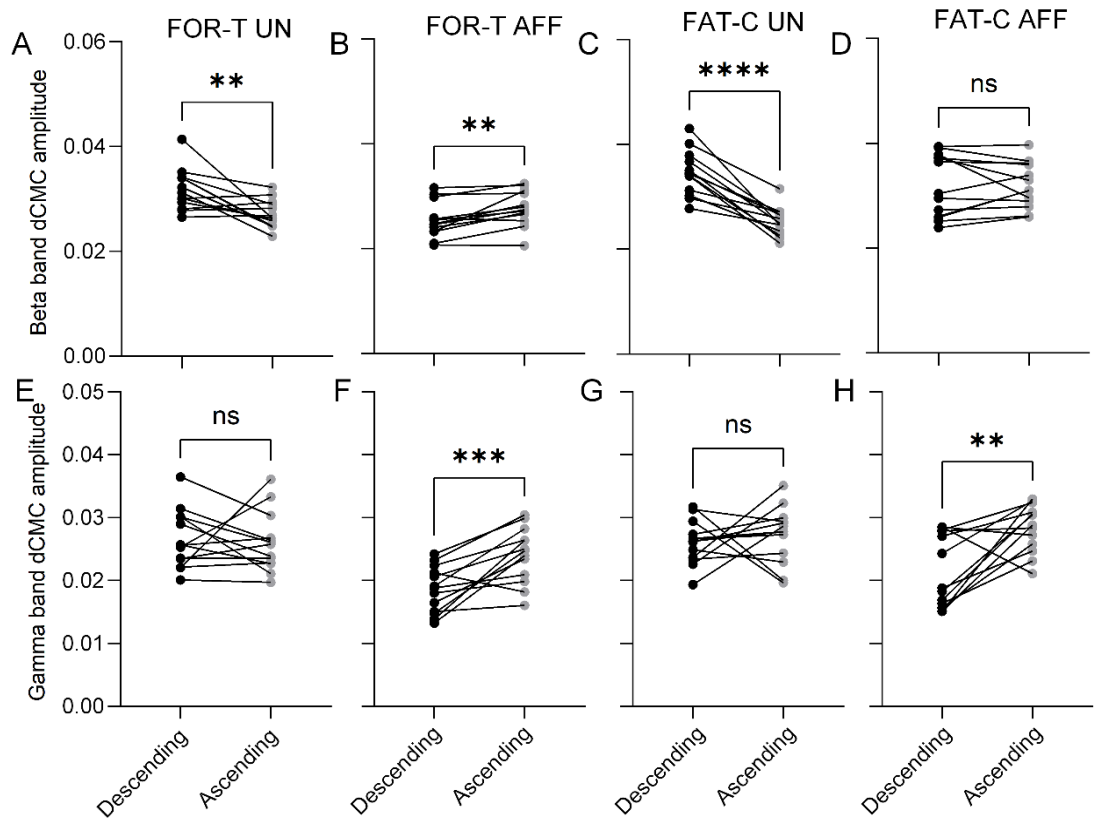

Figure 8 The beta band dCMC dominant main effect in the (A) FOR-T UN group (B) FOR-T AFF group (C) FAT-C UN group (D) FAT-C AFF group and gamma band dCMC dominant main effect in the (E) FOR-T UN group (F) FOR-T AFF group (G) FAT-C UN group (H) FAT-C AFF group. The significant dominant main effect iindicated by "*". The values were represented by mean and SEM at each timepoint.

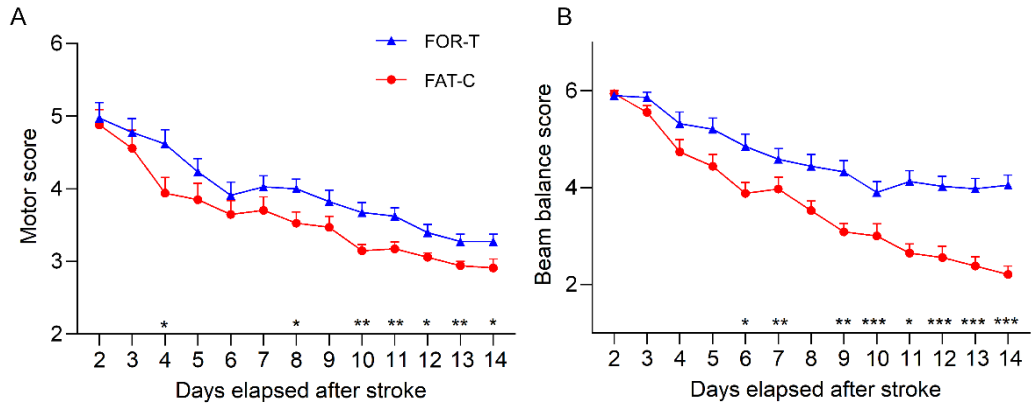

Figure S1 The motor and beam balance scores of the FAT-C and the FOR-T groups from Day2 to Day14 post-stroke. The "*" indicated the significant differences between the FAT-C and FOR-T groups. The values were represented by mean and SEM at each timepoint.

# Reference

Aboy, Mateo, Roberto Hornero, Daniel Abásolo, and Daniel Alvarez. 2006. 'Interpretation of the Lempel-Ziv Complexity Measure in the Context of Biomedical Signal Analysis'. *IEEE Transactions on Bio-Medical Engineering* 53(11):2282–88. doi: 10.1109/TBME.2006.883696.

Amann, Markus, Massimo Venturelli, Stephen J. Ives, John McDaniel, Gwenael Layec, Matthew J. Rossman, and Russell S. Richardson. 2013. 'Peripheral Fatigue Limits Endurance Exercise via a Sensory Feedback-Mediated Reduction in Spinal Motoneuronal Output'. *Journal of Applied Physiology (Bethesda, Md.: 1985)* 115(3):355–64. doi: 10.1152/japplphysiol.00049.2013.

Baker, Stuart N. 2007. 'Oscillatory Interactions between Sensorimotor Cortex and the Periphery'. *Current Opinion in Neurobiology* 17(6):649–55. doi: 10.1016/j.conb.2008.01.007.

Ballester, Belén Rubio, Nick S. Ward, Fran Brander, Martina Maier, Kate Kelly, and Paul F. M. J. Verschure. 2022. 'Relationship between Intensity and Recovery in Post-Stroke Rehabilitation: A Retrospective Analysis'. *Journal of Neurology, Neurosurgery, and Psychiatry* 93(2):226–28. doi: 10.1136/jnnp-2021-326948.

Brasil-Neto, Joaquim P., Leonardo G. Cohen, and Mark Hallett. 1994. 'Central Fatigue as Revealed by Postexercise Decrement of Motor Evoked Potentials'. *Muscle & Nerve* 17(7):713–19. doi: 10.1002/mus.880170702.

Cassidy, Jessica M., Anirudh Wodeyar, Jennifer Wu, Kiranjot Kaur, Ashley K. Masuda, Ramesh Srinivasan, and Steven C. Cramer. 2020. 'Low-Frequency Oscillations Are a Biomarker of Injury and Recovery after Stroke'. *Stroke* 51(5):1442–50.

Casula, Elias Paolo, Maria Concetta Pellicciari, Sonia Bonnì, Barbara Spanò, Viviana Ponzo, Ilenia Salsano, Giovanni Giulietti, Alex Martino Cinnera, Michele Maiella, Ilaria Borghi, Lorenzo Rocchi, Marco Bozzali, Fabrizio Sallustio, Carlo Caltagirone, and Giacomo Koch. 2021. 'Evidence for Interhemispheric Imbalance in Stroke Patients as Revealed by Combining Transcranial Magnetic Stimulation and Electroencephalography'. *Human Brain Mapping* 42(5):1343–58. doi: 10.1002/hbm.25297.

Chen, Chi-Chun, Yu-Lin Wang, and Ching-Ping Chang. 2019. 'Remarkable Cell Recovery from Cerebral Ischemia in Rats Using an Adaptive Escalator-Based Rehabilitation Mechanism'. *PLoS One* 14(10):e0223820.

Chen, Songmei, Xiaolin Zhang, Xixi Chen, Zhiqing Zhou, Weiqin Cong, KaYee Chong, Qing Xu, Jiali Wu, Zhaoyuan Li, Wanlong Lin, and Chunlei Shan. 2023. 'The Assessment of Interhemispheric Imbalance Using Functional Near-Infrared Spectroscopic and Transcranial Magnetic Stimulation for Predicting Motor Outcome after Stroke'. *Frontiers in Neuroscience* 17. doi: 10.3389/fnins.2023.1231693.

Cicinelli, Paola, Patrizio Pasqualetti, Marina Zaccagnini, Raimondo Traversa, Massimiliano Oliveri, and Paolo Maria Rossini. 2003. 'Interhemispheric Asymmetries of Motor Cortex Excitability in the Postacute Stroke Stage: A Paired-Pulse Transcranial Magnetic Stimulation Study'. *Stroke* 34(11):2653–58. doi: 10.1161/01.STR.0000092122.96722.72.

Coleman, Elisheva R., Rohitha Moudgal, Kathryn Lang, Hyacinth I. Hyacinth, Oluwole O. Awosika, Brett M. Kissela, and Wuwei Feng. 2017. 'Early Rehabilitation After Stroke: A Narrative Review'. *Current Atherosclerosis Reports* 19(12):59. doi: 10.1007/s11883-017-0686-6.

Demirtas-Tatlidede, Asli, Miguel Alonso-Alonso, Ravi P. Shetty, Itamar Ronen, Alvaro Pascual-Leone, and Felipe Fregni. 2015. 'Long-Term Effects of Contralesional rTMS in Severe Stroke: Safety, Cortical Excitability, and Relationship with Transcallosal Motor Fibers'. *NeuroRehabilitation* 36(1):51–59.

Deshpande, Gopikrishna, and Xiaoping Hu. 2011. 'Fatigue-Induced Changes in Brain Nonlinearity Inferred by Nonparametric and Differential Equation Models of fMRI'. Pp. 336–41 in *2011 IEEE International Conference on Industrial Technology*. IEEE.

Di Pino, Giovanni, Giovanni Pellegrino, Giovanni Assenza, Fioravante Capone, Florinda Ferreri, Domenico Formica, Federico Ranieri, Mario Tombini, Ulf Ziemann, and John C. Rothwell. 2014. 'Modulation of Brain Plasticity in Stroke: A Novel Model for Neurorehabilitation'. *Nature Reviews Neurology* 10(10):597–608.

Dimyan, Michael A., and Leonardo G. Cohen. 2011. 'Neuroplasticity in the Context of Motor Rehabilitation after Stroke'. *Nature Reviews Neurology* 7(2):76–85.

Dodd, Keith C., Veena A. Nair, and Vivek Prabhakaran. 2017. 'Role of the Contralesional vs. Ipsilesional Hemisphere in Stroke Recovery'. *Frontiers in Human Neuroscience* 11. doi: 10.3389/fnhum.2017.00469.

Fabero-Garrido, Raúl, Tamara del Corral, Santiago Angulo-Díaz-Parreño, Gustavo Plaza-Manzano, Patricia Martín-Casas, Joshua A. Cleland, César Fernández-de-las-Peñas, and Ibai López-de-Uralde-Villanueva. 2022. 'Respiratory Muscle Training Improves Exercise Tolerance and Respiratory Muscle Function/Structure Post-Stroke at Short Term: A Systematic Review and Meta-Analysis'. *Annals of Physical and Rehabilitation Medicine* 65(5):101596. doi: 10.1016/j.rehab.2021.101596.

Faes, Luca, and Giandomenico Nollo. 2011. 'Multivariate Frequency Domain Analysis of Causal Interactions in Physiological Time Series'. *Biomedical Engineering, Trends in Electronics, Communications and Software* 8:403–28.

Finnigan, Simon, Andrew Wong, and Stephen Read. 2016. 'Defining Abnormal Slow EEG Activity in Acute Ischaemic Stroke: Delta/Alpha Ratio as an Optimal QEEG Index'. *Clinical Neurophysiology* 127(2):1452–59.

Graterol Pérez, José A., Stephanie Guder, Chi-un Choe, Christian Gerloff, and Robert Schulz. 2022. 'Relationship Between Cortical Excitability Changes and Cortical Thickness in Subcortical Chronic Stroke'. *Frontiers in Neurology* 13:802113. doi: 10.3389/fneur.2022.802113.

Gwin, Joseph T., Klaus Gramann, Scott Makeig, and Daniel P. Ferris. 2011. 'Electrocortical Activity Is Coupled to Gait Cycle Phase during Treadmill Walking'. *Neuroimage* 54(2):1289–96.

Hendricks, Henk T., Jacques Van Limbeek, Alexander C. Geurts, and Machiel J. Zwarts. 2002. 'Motor Recovery after Stroke: A Systematic Review of the Literature'. *Archives of Physical Medicine and Rehabilitation* 83(11):1629–37.

Hogan, Patrick S., Steven X. Chen, Wen Wen Teh, and Vikram S. Chib. 2020. 'Neural Mechanisms Underlying the Effects of Physical Fatigue on Effort-Based Choice'. *Nature Communications* 11(1):4026. doi: 10.1038/s41467-020-17855-5.

Hsu, Lin-I., Kai-Wen Lim, Ying-Hui Lai, Chen-Sheng Chen, and Li-Wei Chou. 2023. 'Effects of Muscle Fatigue and Recovery on the Neuromuscular Network after an Intermittent Handgrip Fatigue Task: Spectral Analysis of Electroencephalography and Electromyography Signals'. *Sensors* 23(5):2440. doi: 10.3390/s23052440.

Ibáñez-Molina, Antonio J., Vanessa Lozano, María F. Soriano, José I. Aznarte, Carlos J. Gómez-Ariza, and M. T. Bajo. 2018. 'EEG Multiscale Complexity in Schizophrenia during Picture Naming'. *Frontiers in Physiology* 9:1213.

Joshi, Shubham Khemchand, and Stephen Dando. 2024. 'Effect of Sub-Maximal Physical Fatigue on Auditory and Visual Reaction Time in Healthy Adults: Repeated Measures Design'. *Bulletin of Faculty of Physical Therapy* 29(1):30. doi: 10.1186/s43161-024-00196-5.

Jun Hong. 2012. 'The Mental Workload Judgment in Visual Cognition under Multitask Meter Scheme'. *International Journal of the Physical Sciences* 7(5). doi: 10.5897/IJPS11.1372.

KARAKAWA, Sachise, Rumi NISHIMOTO, Masashi HARADA, Naoko ARASHIDA, and Akira NAKAYAMA. 2019. 'Simultaneous Analysis of Tryptophan and Its Metabolites in Human Plasma Using Liquid Chromatography–Electrospray Ionization Tandem Mass Spectrometry'. *Chromatography* 40(3):127–33.

Khademi, Fatemeh, Georgios Naros, Ali Nicksirat, Dominic Kraus, and Alireza Gharabaghi. 2022. 'Rewiring Cortico-Muscular Control in the Healthy and Poststroke Human Brain with Proprioceptive β-Band Neurofeedback'. *The Journal of Neuroscience* 42(36):6861–77. doi: 10.1523/JNEUROSCI.1530-20.2022.

Kinoshita, Shoji, Hiroaki Tamashiro, Takatsugu Okamoto, Naoki Urushidani, and Masahiro Abo. 2019. 'Association between Imbalance of Cortical Brain Activity and Successful Motor Recovery in Sub-Acute Stroke Patients with Upper Limb Hemiparesis: A Functional

near-Infrared Spectroscopy Study'. *Neuroreport* 30(12):822–27. doi: 10.1097/WNR.0000000000001283.

Kotan, Shinichi, Sho Kojima, Shota Miyaguchi, Kazuhiro Sugawara, and Hideaki Onishi. 2015. 'Depression of Corticomotor Excitability after Muscle Fatigue Induced by Electrical Stimulation and Voluntary Contraction'. *Frontiers in Human Neuroscience* 9:363.

Kuppuswamy, A., E. V. Clark, K. S. Sandhu, J. C. Rothwell, and N. S. Ward. 2015. 'Post-Stroke Fatigue: A Problem of Altered Corticomotor Control?' *Journal of Neurology, Neurosurgery & Psychiatry* 86(8):902–4.

Langhorne, Peter, Fiona Coupar, and Alex Pollock. 2009. 'Motor Recovery after Stroke: A Systematic Review'. *The Lancet Neurology* 8(8):741–54.

Langhorne, Peter, Olivia Wu, Helen Rodgers, Ann Ashburn, and Julie Bernhardt. 2017. 'A Very Early Rehabilitation Trial after Stroke (AVERT): A Phase III, Multicentre, Randomised Controlled Trial'. *Health Technology Assessment (Winchester, England)* 21(54):1–120. doi: 10.3310/hta21540.

Lanzone, J., M. A. Colombo, S. Sarasso, F. Zappasodi, M. Rosanova, M. Massimini, V. Di Lazzaro, and G. Assenza. 2022. 'EEG Spectral Exponent as a Synthetic Index for the Longitudinal Assessment of Stroke Recovery'. *Clinical Neurophysiology* 137:92–101.

Leemburg, Susan, Bo Gao, Ertugrul Cam, Johannes Sarnthein, and Claudio L. Bassetti. 2018. 'Power Spectrum Slope Is Related to Motor Function after Focal Cerebral Ischemia in the Rat'. *Sleep* 41(10):zsy132.

Lefaucheur, Jean-Pascal, André Aleman, Chris Baeken, David H. Benninger, Jérôme Brunelin, Vincenzo Di Lazzaro, Saša R. Filipović, Christian Grefkes, Alkomiet Hasan, Friedhelm C. Hummel, Satu K. Jääskeläinen, Berthold Langguth, Letizia Leocani, Alain Londero, Raffaele Nardone, Jean-Paul Nguyen, Thomas Nyffeler, Albino J. Oliveira-Maia, Antonio Oliviero, Frank Padberg, Ulrich Palm, Walter Paulus, Emmanuel Poulet, Angelo Quartarone, Fady Rachid, Irena Rektorová, Simone Rossi, Hanna Sahlsten, Martin Schecklmann, David Szekely, and Ulf Ziemann. 2020. 'Evidence-Based Guidelines on the Therapeutic Use of Repetitive Transcranial Magnetic Stimulation (rTMS): An Update (2014-2018)'. *Clinical Neurophysiology: Official Journal of the International Federation of Clinical Neurophysiology* 131(2):474–528. doi: 10.1016/j.clinph.2019.11.002.

Li, Bo, Sican Liu, Dingyin Hu, Guanghui Li, Rongyu Tang, Da Song, Yiran Lang, and Jiping He. 2021. 'Electrocortical Activity in Freely Walking Rats Varies with Environmental Conditions'. *Brain Research* 1751:147188.

Li, Fengwu, John T. Pendy Jr, Jessie N. Ding, Changya Peng, Xiaorong Li, Jiamei Shen, Sainan Wang, and Xiaokun Geng. 2017. 'Exercise Rehabilitation Immediately Following Ischemic Stroke Exacerbates Inflammatory Injury'. *Neurological Research*.

Li, Jiali, Hewei Wang, Yujian Yuan, Yunhui Fan, Fan Liu, Jingjing Zhu, Qing Xu, Lan Chen, Miao Guo, Zhaoying Ji, Yun Chen, Qiurong Yu, Tianhao Gao, Yan Hua, Mingxia Fan, and Limin Sun. 2022. 'Effects of High Frequency rTMS of Contralesional Dorsal Premotor Cortex in Severe Subcortical Chronic Stroke: Protocol of a Randomized Controlled Trial with Multimodal Neuroimaging Assessments'. *BMC Neurology* 22(1):125. doi: 10.1186/s12883-022-02629-x.

Li, Wang. 2022. 'Evaluation of Running Intensity and Fatigue Degree Based on Human Physiological Information' edited by H. A. Khattak. *Mobile Information Systems* 2022:1–7. doi: 10.1155/2022/6488312.

Liang, Tie, Qingyu Zhang, Xiaoguang Liu, Bin Dong, Xiuling Liu, and Hongrui Wang. 2021. 'Identifying Bidirectional Total and Non-Linear Information Flow in Functional Corticomuscular Coupling during a Dorsiflexion Task: A Pilot Study'. *Journal of NeuroEngineering and Rehabilitation* 18:1–15.

Liu, Jinbiao, Yixuan Sheng, and Honghai Liu. 2019. 'Corticomuscular Coherence and Its Applications: A Review'. *Frontiers in Human Neuroscience* 13:100. doi: 10.3389/fnhum.2019.00100.

Luong, Tinh N., Holly J. Carlisle, Amber Southwell, and Paul H. Patterson. 2011. 'Assessment of Motor Balance and Coordination in Mice Using the Balance Beam'. *Journal of Visualized Experiments: JoVE* (49):2376.

Mahrukh, Bushra Riaz, Zarqa Sharif, Usama Mahmood, Hanan Azfar, and Muhammad Junaid Ilyas. 2023. 'Efficacy of High-Intensity Interval Training versus Moderate-Intensity Continuous Training in Chronic Stroke Rehabilitation'. *Journal of Health and Rehabilitation Research* 3(2):187–93. doi: 10.61919/jhrr.v3i2.106.

Mauro, M. C., A. Fasano, M. Germanotta, L. Cortellini, S. Insalaco, A. Pavan, A. Comanducci, E. Guglielmelli, and I. Aprile. 2024. 'Restoring of Interhemispheric Symmetry in Patients with Stroke Following Bilateral or Unilateral Robot-Assisted Upper-Limb Rehabilitation: A Pilot Randomized Controlled Trial'. *IEEE Transactions on Neural Systems and Rehabilitation Engineering*.

McKay, W. Barry, Stephen M. Tuel, Arthur M. Sherwood, Dobrivoje S. Stokić, and Milan R. Dimitrijević. 1995. 'Focal Depression of Cortical Excitability Induced by Fatiguing Muscle Contraction: A Transcranial Magnetic Stimulation Study'. *Experimental Brain Research* 105(2):276–82. doi: 10.1007/BF00240963.

Meeusen, Romain, Philip Watson, Hiroshi Hasegawa, Bart Roelands, and Maria F. Piacentini. 2006. 'Central Fatigue: The Serotonin Hypothesis and Beyond'. *Sports Medicine* 36(10):881–909. doi: 10.2165/00007256-200636100-00006.

Myers, Christopher. 2024. 'Central and Peripheral Fatigue'. Pp. 293–320 in *Skeletal Muscle Physiology: An Update to Anatomy and Function*, edited by C. Myers. Cham: Springer

Nature Switzerland.

Persico, Antonio M., Elisa Mengual, Rainald Moessner, Scott F. Hall, Randal S. Revay, Ichiro Sora, Jon Arellano, Javier DeFelipe, José Manuel Giménez-Amaya, and Monica Conciatori. 2001. 'Barrel Pattern Formation Requires Serotonin Uptake by Thalamocortical Afferents, and Not Vesicular Monoamine Release'. *Journal of Neuroscience* 21(17):6862–73.

Reddin, Catriona, Robert Murphy, Graeme J. Hankey, Conor Judge, Denis Xavier, Annika Rosengren, John Ferguson, Alberto Alvarez-Iglesias, Shahram Oveisgharan, Helle K. Iversen, Fernando Lanas, Fawaz Al-Hussein, Anna Czlonkowska, Aytekin Oguz, Clodagh McDermott, Nana Pogosova, German Málaga, Peter Langhorne, Xingyu Wang, Mohammad Wasay, Salim Yusuf, Martin O'Donnell, and INTERSTROKE investigators. 2022. 'Association of Psychosocial Stress With Risk of Acute Stroke'. *JAMA Network Open* 5(12):e2244836. doi: 10.1001/jamanetworkopen.2022.44836.

Schaar, Krystal L., Miranda M. Brenneman, and Sean I. Savitz. 2010. 'Functional Assessments in the Rodent Stroke Model'. *Experimental & Translational Stroke Medicine* 2(1):13. doi: 10.1186/2040-7378-2-13.

Stuart, Douglas G., and Hans Hultborn. 2008. 'Thomas Graham Brown (1882–1965), Anders Lundberg (1920–), and the Neural Control of Stepping'. *Brain Research Reviews* 59(1):74–95.

Sun, Jing, Zheng Ke, Shea Ping Yip, Xiao-ling Hu, Xiao-xiang Zheng, and Kai-yu Tong. 2014. 'Gradually Increased Training Intensity Benefits Rehabilitation Outcome after Stroke by BDNF Upregulation and Stress Suppression'. *BioMed Research International* 2014:1–8. doi: 10.1155/2014/925762.

Tam, Pui Kit, Nicodemus Edrick Oey, Ning Tang, Guhan Ramamurthy, and Effie Chew. 2024. 'Facilitating Corticomotor Excitability of the Contralesional Hemisphere Using Non-Invasive Brain Stimulation to Improve Upper Limb Motor Recovery from Stroke—A Scoping Review'. *Journal of Clinical Medicine* 13(15):4420. doi: 10.3390/jcm13154420.

Tang, Qing, Guangming Li, Tao Liu, Anguo Wang, Shenggang Feng, Xiang Liao, Yu Jin, Zhiwei Guo, Bin He, Morgan A. McClure, Guoqiang Xing, and Qiwen Mu. 2015. 'Modulation of Interhemispheric Activation Balance in Motor-Related Areas of Stroke Patients with Motor Recovery: Systematic Review and Meta-Analysis of fMRI Studies'. *Neuroscience and Biobehavioral Reviews* 57:392–400. doi: 10.1016/j.neubiorev.2015.09.003.

Taylor, Janet L., Markus Amann, Jacques Duchateau, Romain Meeusen, and Charles L. Rice. 2016. 'Neural Contributions to Muscle Fatigue: From the Brain to the Muscle and Back Again'. *Medicine and Science in Sports and Exercise* 48(11):2294–2306. doi: 10.1249/MSS.0000000000000923.

Tergau, Frithjof, Rolf Geese, Axel Bauer, Susanne Baur, WALTER PAULUS, and Carl Detlev Reimers. 2000. 'Motor Cortex Fatigue in Sports Measured by Transcranial Magnetic Double Stimulation'. *Medicine & Science in Sports & Exercise* 32(11):1942–48.

Tornero-Aguilera, José Francisco, Jorge Jimenez-Morcillo, Alejandro Rubio-Zarapuz, and Vicente J. Clemente-Suárez. 2022. 'Central and Peripheral Fatigue in Physical Exercise Explained: A Narrative Review'. *International Journal of Environmental Research and Public Health* 19(7):3909.

Van Wijngaarden, Joeri BG, Riccardo Zucca, Simon Finnigan, and Paul FMJ Verschure. 2016. 'The Impact of Cortical Lesions on Thalamo-Cortical Network Dynamics after Acute Ischaemic Stroke: A Combined Experimental and Theoretical Study'. *PLoS Computational Biology* 12(8):e1005048.

Wan, Jing-jing, Zhen Qin, Peng-yuan Wang, Yang Sun, and Xia Liu. 2017. 'Muscle Fatigue: General Understanding and Treatment'. *Experimental & Molecular Medicine* 49(10):e384–e384.

Wang, Dejuan, Xiaojie Wang, Penglai Liu, Siqi Jing, Han Du, Lingzhi Zhang, Fan Jia, and Anan Li. 2020. 'Serotonergic Afferents from the Dorsal Raphe Decrease the Excitability of Pyramidal Neurons in the Anterior Piriform Cortex'. *Proceedings of the National Academy of Sciences* 117(6):3239–47. doi: 10.1073/pnas.1913922117.

Wang, Qi, Dai Zhang, Ying-Yu Zhao, Hong Hai, and Yue-Wen Ma. 2020. 'Effects of High-Frequency Repetitive Transcranial Magnetic Stimulation over the Contralesional Motor Cortex on Motor Recovery in Severe Hemiplegic Stroke: A Randomized Clinical Trial'. *Brain Stimulation* 13(4):979–86. doi: 10.1016/j.brs.2020.03.020.

Wang, Wei, Ming Wei, Yuanyuan Cheng, Hua Zhao, Hutao Du, Weijia Hou, Yang Yu, Zhizhong Zhu, Lina Qiu, Tao Zhang, and Jialing Wu. 2022. 'Safety and Efficacy of Early Rehabilitation After Stroke Using Mechanical Thrombectomy: A Pilot Randomized Controlled Trial'. *Frontiers in Neurology* 13. doi: 10.3389/fneur.2022.698439.

Wei, Xufang, Shengtong Sun, Manyu Zhang, and Zhenqiang Zhao. 2024. 'A Systematic Review and Meta-Analysis of Clinical Efficacy of Early and Late Rehabilitation Interventions for Ischemic Stroke'. *BMC Neurology* 24(1):91. doi: 10.1186/s12883-024-03565-8.

Xu, Rui, Chuncui Zhang, Feng He, Xin Zhao, Hongzhi Qi, Peng Zhou, Lixin Zhang, and Dong Ming. 2018. 'How Physical Activities Affect Mental Fatigue Based on EEG Energy, Connectivity, and Complexity'. *Frontiers in Neurology* 9:915. doi: 10.3389/fneur.2018.00915.

Xu, Yuchen, Yuanfa Yao, Hao Lyu, Stephanie Ng, Yingke Xu, Wai Sang Poon, Yongping Zheng, Shaomin Zhang, and Xiaoling Hu. 2020. 'Rehabilitation Effects of Fatigue-Controlled Treadmill Training after Stroke: A Rat Model Study'. *Frontiers in Bioengineering and Biotechnology* 8:590013.

Zeng, Hongji, Jiaying Yang, Junfa Wu, Yu Ding, Shuya Yuan, Rui Wang, Weijia Zhao, and Xi Zeng. 2024. 'The Impact of Post-Stroke Fatigue on Inpatient Rehabilitation Outcomes: An Observational Study'. *PLOS ONE* 19(5):e0302574. doi: 10.1371/journal.pone.0302574.

Zhang, Lei, Jingwen Liu, and Mingsheng Liu. 2024. 'Transsynaptic Degeneration of Ventral Horn Motor Neurons Exists but Plays a Minor Role in Lower Motor System Dysfunction in Acute Ischemic Rats'. *PLOS ONE* 19(4):e0298006. doi: 10.1371/journal.pone.0298006.

Zhou, Sa, Ziqi Guo, Kiufung Wong, Hanlin Zhu, Yanhuan Huang, Xiaoling Hu, and Yong-Ping Zheng. 2021. 'Pathway-Specific Cortico-Muscular Coherence in Proximal-to-Distal Compensation during Fine Motor Control of Finger Extension after Stroke'. *Journal of Neural Engineering* 18(5):056034.